\documentclass[12pt,preprint]{aastex}

\slugcomment{to be submitted to the Astrophysical Journal.}

\shorttitle{HST observations and modeling of  PKS 0637-752}
\shortauthors{Mehta, Georganopoulos, Perlman, Padgett, Chartas}

\begin{document}

\title{ HST observations of the quasar PKS 0637-752: equipartition electron-proton jet from the most complete spectral coverage to date}

\author{Kushal T. Mehta\altaffilmark{1},
Markos Georganopoulos\altaffilmark{1,2},
Eric S. Perlman \altaffilmark{3}, 
Charles A. Padgett\altaffilmark{1,2}
George Chartas \altaffilmark{4}}

\altaffiltext{1}{Department of Physics, 
University of Maryland, Baltimore County, 
1000 Hilltop Circle, Baltimore, MD 21250, USA.}
\altaffiltext{2}{NASA Goddard Space Flight Center, Code 663, Greenbelt, MD 20771, USA.}
\altaffiltext{3}{Department of Physics and Space Sciences, 
Florida Insitute of Technology, 
150 West University Boulevard, Melbourne, FL  32901, USA.}
\altaffiltext{4}{Department of Astronomy and Astrophysics,
Penn State University
525 Davey Lab, University Park, PA 16802 , USA.}

\begin{abstract}
We present new NICMOS and ACS observations of the quasar jet PKS 0637-752, and we use them, together with existing multiwavelength observations, to produce the most  complete spectral coverage of the source to date. We explore the implications of these observations in the context of models for the jet X-ray emission. 
By relaxing the assumption of equipartition, we undertake an exhaustive  study of the parameter space for external Compton off the CMB  (EC/CMB) model.
We find that  the multiwavelength observations  exclude a magnetic field dominated jet.  Using the method proposed by Georganopoulos et al. (2005) for probing the jet matter content 
we show that protons are needed for practically all jet configurations, extending a previous application of the method by Uchiyama et al. (2005) that was based on exploring  three particular jet configurations.
  We also show that equipartition is the only configuration that can reproduce the observations and have one proton per radiating lepton. 
 We finally present a rather model - independent argument that the jet has a spine-sheath flow pattern, with the spine being faster and emitting most of the IR-optical-X-ray emission.

\end{abstract}


\section{Introduction \label{intro}}

% The saga of PKS 0637 and the failure of SSC.

When the \emph{Chandra X-Ray Observatory} used the bright X-ray quasar PKS 0637-752 to  focus  its mirror assembly during its first observation, astronomers expected to detect  a  bright X-ray point source. 
However, in addition to the bright core of the quasar, \emph{Chandra}  detected significant X-ray emission from the previously detected  quasar radio jet   \citep{schwartz00,chartas00}. This bright X-ray emission was found to overproduce the X-ray flux that corresponds to the extension of the radio-optical spectrum to X-ray energies,
arguing for a second spectral component, for which the most plausible candidate was initially thought to be  synchrotron - self Compton (SSC) emission. However,  it became immediately apparent  from modeling the  multi-wavelength spectral energy distribution (SED) of the knots of  PKS 0637-752 that  the  expected SSC X-ray flux in equipartition conditions
 severely underpredicts the observed X-ray flux, with the case for SSC only getting  worse when relativistic beaming effects  are considered \citep{schwartz00,chartas00}. 
 
 % Other CXO observations
 
 Following the discovery of the X-ray jet of PKS 0637-752, there has been extensive observational and 
theoretical work on powerful quasar jets (for a recent review see Harris \& Krawczynski 2006). Further X-ray studies using \emph{Chandra} have confirmed that bright X-ray emitting jets are common among radio-loud quasars and radio galaxies  (e.g. Sambruna et al. 2002, 2004, 2006; Marshall et al. 2005; Siemiginowska et al. 2002; Kataoka et al. 2003; Jorstad et al. 2004; Cheung, Stawarz, \& Siemiginowska 2006;  Schwartz et al. 2006). They also confirmed that in these jets the X-ray emission is a spectral component separate from the radio optical synchrotron emission, with the X-ray flux being higher that that anticipated by simply extending the radio-optical spectrum to X-ray energies. 

% The EC/CMB model and its baggage  

A plausible candidate for the X-ray emission mechanism was suggested by 
 \cite{tavecchio00} and \cite{celotti01}. These authors argued that the X-ray flux of the knots of PKS 0637-752 is due to external Compton emission  from jet relativistic electrons  in equipartition with the magnetic field of the jet,  up-scattering  cosmic microwave background photons (EC/CMB).   The model, modivated by the superluminal motions observed in  the sub-pc scale jet \citep{lovell00},  requires that (i) 
the large scale jet is significantly beamed (bulk Lorentz factor $\Gamma\sim 10$) and (ii) 
the electron distribution has a low energy break or cutoff at $\gamma \lesssim 100 $. Both these requirements increase the required jet power uncomfortably close to the Eddington limit \citep{dermer04}.
Because in the EC/CMB model  the X-rays  are produced by very low energy electrons ($\gamma \sim 100$), these electrons  have long cooling times that  allow them to reach the terminal points of even megaparsec scale  jets without losing a significant fraction of their energy.
Thus,  in the EC/CMB we would expect to see continuous X-ray emission throughout the jet instead of the  knotty X-ray structure we usually observe. A possible solution to this issue is to assume that jets are not continuous and that the X-ray knots are ejected from the central engine during periods of high activity and retain their knot structure as they propagate downstream.  Also, in several X-ray jets, the ratio of the radio to X-ray flux in the knots decreases as we move away from the core. In the context of the EC/CMB model this can be explained if jets gradually decelerate \citep{georganopoulos04}. 
 In some jets, including PKS 0637-752, there is a low emission bridge of $\sim 5-10$ arcsec connecting the quasar core to the first jet knot.
 \cite{georganopoulos05} argued that, in the context of the EC/CMB model, this can be used to constrain the matter content of the jet through IR observations of the bridge, and \cite{uchiyama05} used this to show, through {\sl Spitzer} observations of PKS 0637-752, that an electron-positron jet is not favored  if the X-rays are due to the EC/CMB mechanism.
 
 %Synchrotron X-rays anybody?
 
 The fact that  the EC/CMB model requires bulk Lorentz factors much in excess than required by jet-to-counter-jet radio asymmetry arguments (e.g. Arshakian \& Longair 2004), together with the  issues just discussed, raises the possibility that the X-rays are of synchrotron nature (e.g. review by Harris \& Krawczynski 2006) by an additional population
 of very high energy ($\sim 100$ TeV) electrons. Such high energy electrons may be produced
 if the jet is characterized by velocity shear \citep{stawarz02, rieger04}. Alternatively, the decay of a collimated neutron beam produced in the sub-pc scale jet can provide a second high energy electron
 component \citep{atoyan04}.
 The synchrotron  option is viable for sources such as 3C 273, for which the high energy component extends smoothly down to optical energies  \citep{uchiyama06, jester06, jester07}, but runs into difficulties for sources like PKS 0637-752, in which the high energy component cuts off before the optical regime, because these high energy electrons would cool rapidly to 
lower energies and would  radiate in the optical,  producing, thereby,  a much higher optical flux than  observed. 
If, indeed, the X-ray emission is of synchrotron nature, the multi TeV electrons responsible for it
will unavoidably up-scatter the CMB photons to TeV energies \citep{georganopoulos06}. For 3C 273, the upper limit from shallow HESS TeV observations is compatible with the synchrotron interpretation, provided $\delta \lesssim 10 $, where $\delta $ is the usual Doppler factor.

Here we present HST NICMOS and ACS observations of PKS 0637-752, and discuss the constraints they pose, together with available multiwavelength observations,  on our understanding of the jet physics.  In \S \ref{obs} we present  the data reduction procedure, in \S \ref{results} our observational results, together with existing multiwavelength observations,    in \S \ref{discussion}  the  constraints on the jet physics, and in \S \ref{conclusions} our conclusions.

\section{Observations and data reduction\label{obs}}

PKS 0637-752 ($z=0.651$, luminosity distance $d_L=3895$ Mpc) has been observed in different energy bands by other researchers. Here we use the  8.6 GHz radio observations taken with  the Australian Telescope Compact Array  (Lovell et al. 2000), the Spitzer observations by Uchiyama et al. (2005),  the HST WFPC 2 observations by Schwartz et al. (2000), and the Chandra X-ray observations  by Chartas et al. (2000) and Schwartz et al. (2000).

%We start with the optical (NICMOS and ACS) with data reduction process:

Our HST observations took place  on November 15th and 16th 2005 using the ACS and NICMOS instruments respectively.  The data were processed by the standard ``on-the-fly'' reprocessing calibration pipeline. 
We performed subsequent data reduction using the NOAO Image Reduction Analysis Facility (IRAF) and Python 
IRAF (PyRAF) packages.
%ACS
The ACS data was taken using filter F475W and the Wide Field Channel  (WFC) detector which has a pivot wavelength of 
0.4744 $\mu$m with a field of view of 202'' x 202'',  corresponding to a resolution of 0.05 arcsecond/pixel. The 
total exposure time was 2556s. This data was reduced using the standard ``multidrizzle'' script  which completes all 
the data reduction and dithering tasks. These tasks include: creating 
a bad pixel map, sky subtraction, drizzling the data onto separate output images, combining the output images into 
a median image, blotting or undrizzling the combined image, creating a cosmic ray mask by comparing the blotted 
image to the original images and drizzling all the images onto a final mosaic using the bad pixel and cosmic ray 
masks.

%NICMOS

The NICMOS data was taken using the NIC3 camera using filter F160W with a pivot wavelength of 1.604 $\mu$m. The field of 
view of NIC3 camera is 51.2'' x 51.2'' with a resolution of 0.2''/pixel. Although it has a lower angular resolution,  we used the NIC3 camera instead of the NIC1 or 
NIC2 as the NIC3 camera has larger pixels  which would help us detect any faint, extended emission 
from the cold electrons in the jet. The NICMOS observations were split into 3 orbits 
each with an exposure time of 3072s. Orbits 2 and 3 were shifted by 5 and 10 pixels respectively. This enables us to 
detect hot pixels and other CCD defects easily. In the NICMOS data, we come across some important anomalies such as the 
pedestal effect and bad pixels which are not corrected for by multidrizzle.
The pedestal effect is corrected for by ``pedsub'', hot pixels were replaced by the average value of the surrounding pixels 
using the ``imedit'' task. We could not use the ``multidrizzle'' task as it does not account for the pedestal effect and tended 
to oversubtract the sky for the NICMOS data. The compensation was done using tasks from the "dither" package in IRAF. These 
tasks included sky subtraction, drizzling the data with corrections for geometric and other distortions by the CCD, correcting 
for the shifts in images from the 3 orbits and rotating the image to show North up and East left.

\section{Results \label{results}}

%What we see

\begin{figure}
\epsscale{0.6}
\plotone{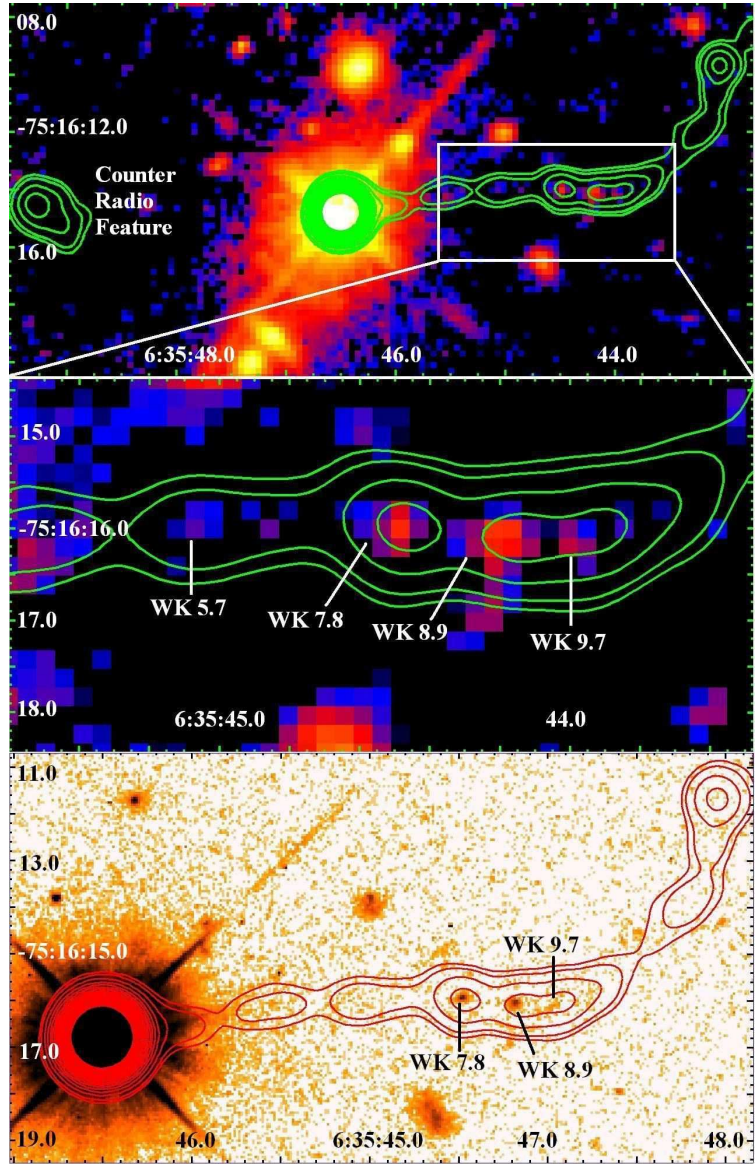}
\caption{The NICMOS and ACS images with the   radio contours for PKS 0637-752 overlaid. 
The top panel is the NICMOS image of the whole jet with green radio contours. The middle panel is a zoomed image of the white box in the top panel. The bottom panel shows the ACS image with radio contours in red.}
\label{fig:maps}
\end{figure}

	The radio jet (8.6 GHz, taken with the Australian Telescope Compact Array; Lovell et al. 2000), shown by the  contours in Figure \ref{fig:maps}, shows a bright jet extending to $\sim 10''$ from the quasar core, 
then bending North-West.
% and eventually ending in a terminal hotspot. 
There is also a counter jet radio feature
approximately $8''-9''$ east of  the core and  at least 2 knots between $7''$ and $10''$ west of  the core with possible 
features further in.  The \emph{Chandra} X-ray image  \citep{chartas00, schwartz00} shows  at least 3 X-ray knots, 
WK 5.7, WK 7.8, WK 8.9 and possibly a fourth knot WK 9.7. 
Previous HST observations (WFPC2, F702W) at an effective wavelength $0.697\, \mu$m \citep{schwartz00} detect the three knots WK 5.7, WK 7.8, WK 8.9. There was also a deeper by a factor of  $\sim$ 2 STIS HST observation (PI C. M. Urry, cycle 10) practically at the same effective wavelength. 
The infrared (\emph{Spitzer}) jet \citep{uchiyama05} is aligned well with the X-ray and radio jet and shows 2 bright knots: WK 7.8 and WK 8.9.
However, the resolution of  \emph{Spitzer} makes it difficult to distinguish these knots clearly. There is no
evidence of the jet after the bend or the counter jet radio feature  in the \emph{Spitzer} and \emph{Chandra} data.

From both the ACS and NICMOS images, we see the bright quasar core and the three bright knots WK 7.8, WK 8.9 and WK 9.7 
separately, with WK 8.9 being the brightest knot. In the NICMOS image, we also see the  inner knot WK 5.7,  which 
was  only seen in X-rays before, and we obtain 
limits from NICMOS and ACS for the counter jet radio feature. Due to the lower spatial resolution of the NIC3 camera, there 
is contamination from a background galaxy  from the South in the WK 8.9 knot (this galaxy  is also seen on the ACS image). The host galaxy contamination is accounted for by 
subtracting the galaxy flux scaled by the ratio of the two areas from the raw flux. The higher resolution of the ACS camera 
helps prevent such contamination in the ACS image.

	 The optical jet is 
well aligned with the radio, X-ray, and infrared jet. We do not  detect  the jet after the bend and the counter jet radio feature  in both the ACS and NICMOS images. With ACS we see that WK 7.8 and WK 8.9 are well confined bright knots whereas WK 9.7 
seems to be extended in the North-West direction. We also detect with NICMOS  a perviously undetected feature along the jet between WK 5.7 and WK 7.8. This 
new feature (WK 6.3) is approximately $6.''3$ from the core and is the least bright feature from all the knots or features we detect. 
Both WK 5.7 and WK 6.3 are enveloped by the same radio feature, and could not have been resolved by the Spitzer observations. For this reason, and to  allow direct comparison  with Uchiyama et al. (2007), in the rest of the paper we use the same size they used for the WK 5.7 knot, which includes the WK 6.3 feature.
 It is interesting to note that even though the brightest pixels in NICMOS do fall within the radio knots, there are parts of knots that lie slightly outside the contours. This may suggest a slight shift in the overall knot position between the optical and radio knots. The higher resolution ACS knots, however, fall well within the radio contours.

\begin{figure}
\epsscale{0.9}
\plotone{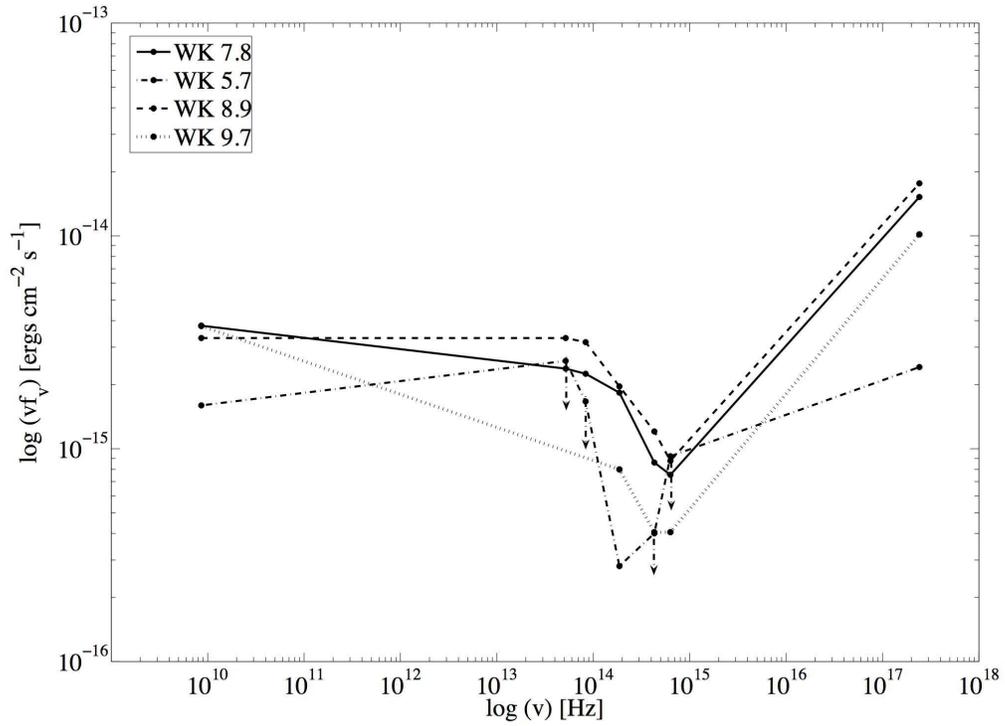}
\caption{The SEDs of the four knots seen in the western jet for PKS 0637-752. The arrows represent upper limits in the SED for knot WK 5.7. The connecting lines are drawn to guide the eye.} 
\label{fig:f_knot_seds}
\end{figure}

\begin{figure}
\epsscale{0.7}
\plotone{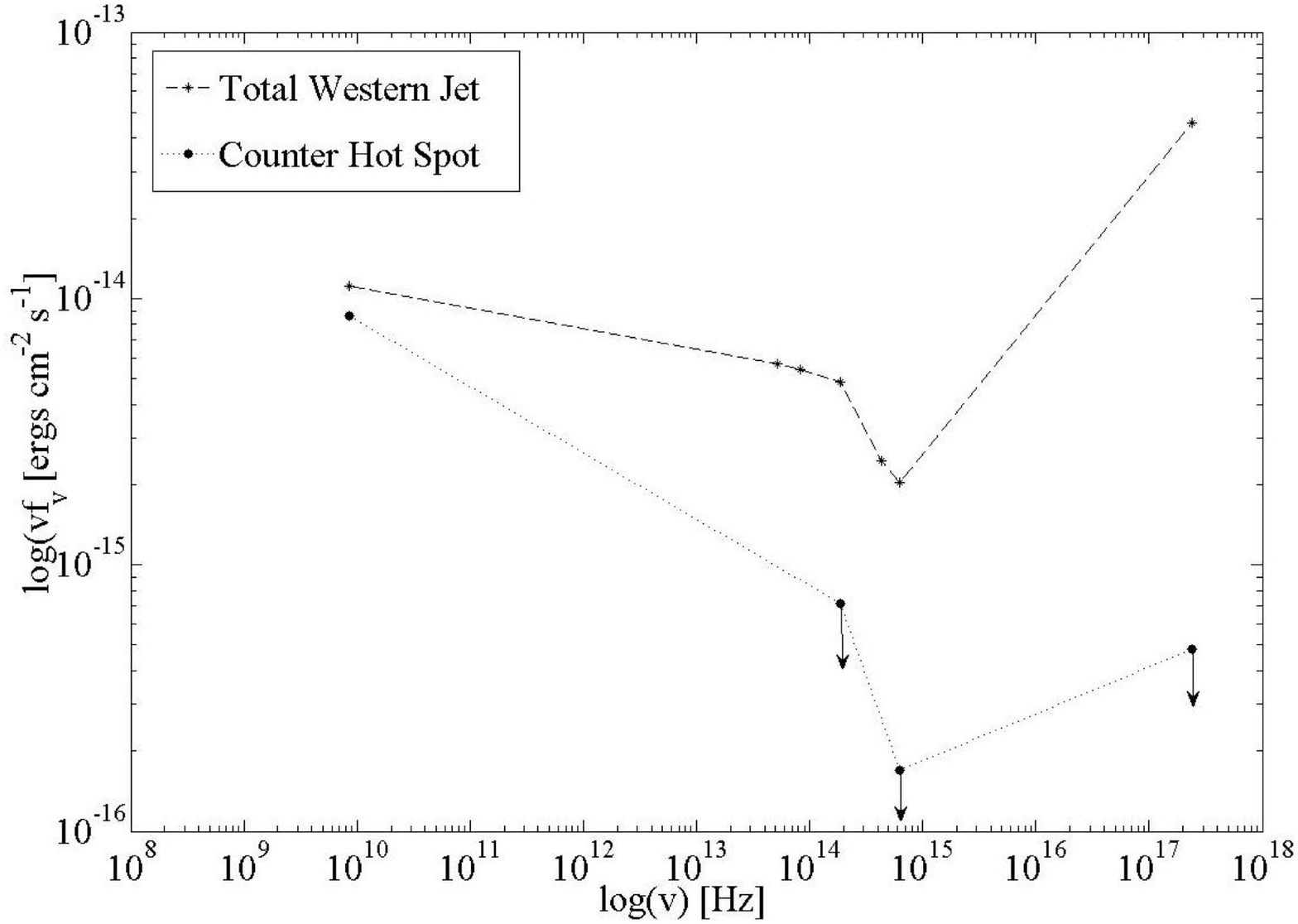}
\caption{  A comparison between the SED of the total western jet (WK 7.8 + WK 8.9 + WK 9.7) to the total eastern jet, noted as the counter radio feature in Figure 1. Note that  while the radio fluxes are comparable, there are only upper limits for the optical and X-ray emission of the eastern jet. As we argue in \S \ref{spine} this suggests a spine-sheath flow with the spine being faster and emitting most of the IR to X-ray radiation. } 
\label{fig:counterjet}
\end{figure}

	The SEDs for the four knots (WK 5.7, 7.8, 8.9 and 9.7) are shown in Figure \ref{fig:f_knot_seds}. We see that all the 
knots have the same basic shape of dropping significantly in the optical before brightening in the X-rays. The three near IR - optical points (NICMOS, WFPC2 and ACS) are interesting as they are the interface between the radio to optical component and the optical to X-rays component of the SED.
An interesting question is if in the tail of the optical component one can discern the rise of the
new component that eventually produces the X-rays.
 For knot 8.9, there is no evidence for this,  as we see the point-to-point 
slopes between the three points drop. Knot WK 7.8 shows an interesting trend near the ACS point. The point to point slopes between the 
NICMOS-WFPC2 points and the WFPC2-ACS points are beyond $3\sigma$ of each other: $\alpha_{NICMOS-WFPC2} = -0.907 \pm 0.055$ and $\alpha_{WFPC2-ACS} 
= -0.345 \pm 0.119$. However, the probability of the three points falling on a line, using the $\chi^2$ test, is 0.15. Thus, 
Knot 7.8 shows suggestive, but not conclusive evidence of an additional component in the ACS frequency range. Knot 9.7 clearly shows 
the presence of an additional component as the ACS flux is approximately the same as the WFPC2 flux. Interestingly, we see that the brightest 
knot shows the least evidence for this component while the least bright knot gives a strong evidence for the same. The SED for the total western 
jet and the SED for the counter jet radio feature  are shown in Figure \ref{fig:counterjet}. The flux of the total jet is calculated by adding up all the 
definitive fluxes from the 4 knots (WK 5.7, 7.8, 8.9, 9.7). The fluxes with limits were ignored so as to obtain a lower limit on the 
total jet flux. The complete list of fluxes is given in Table \ref{tab:fluxes}.

\section{Constraints on the jet physics \label{discussion}}

\subsection{The synchrotron X-ray interpretation \label{sub:synchrotron}}

A possibility  mentioned in \S \ref{intro}  is that the X-ray emission for knots 
with a broadband SED similar to that of knot WK 7.8, in which the X-rays are part of a separate spectral component,  is of synchrotron nature  \citep{hardcastle04, harris06, jester07}. 
A typical one  zone synchrotron model in which the electron distribution is regulated by injection, radiative losses, and escape runs into difficulties because the X-ray emitting electrons have a cooling time significantly faster than their escape time. The observed X-ray spectral index $\alpha_X\approx 0.7$ requires an electron energy distribution (EED) that locally is a power law
$n(\gamma)\propto \gamma^{-p}$, with $p=2\alpha_X+1=2.4$. Because these multi-TeV energy electrons are in the fast cooling regime, the acceleration mechanism that produced them is required to provide an EED $n(\gamma)\propto \gamma^{-1.4}$. Such hard EEDs are far from the usual  2-2.3 index of shock acceleration \citep{kirk00}, although an index of $1.4$ is within the wide range anticipated  in different acceleration scenaria (e.g. Ostrowski 2008).  A more serious issue with one zone models is the 
fact that the level of the observed optical-UV  emission is very low compared to the extrapolation of the X-ray emission in the optical. This is not expected, because radiative cooling should have extended the high energy electron population to lower energies, and these lower energy electrons would fill with their synchrotron emission the optical-UV ``valley''. 
Two-zone synchrotron models can be devised, at the cost of practically doubling the free parameters.
In two-zone models where the acceleration of the X-ray emitting particles takes place in a separate
zone, the optical ``valley'' forces us to adopt an ad-hoc injection of electrons at $\sim 10 $ TeV  energies
which are then accelerated up to at least $\sim 100 $ TeV  before they escape.  This population of electrons must then escape to an environment of much lower magnetic field, so that these electrons
will not produce substantial optical-UV synchrotron emission as they cool.

\subsection{ Bulk Compton constraints on the cold lepton power  \label{bridge}} 

As proposed by \cite{georganopoulos05}, the fact that the  IR emission of  the  `bridge', the 
part of the jet interior to knot WK 7.8,  is very low can be used to derive constraints for the 
lepton power carried by the jet, under the assumption that the jet power in the bridge is carried by some combination of protons and leptons and that the jet flow is not episodic.  
The idea behind the bridge diagnostic is that even if the electrons are cold ($\gamma\approx 1$) in the bulk flow frame, they will up-scatter the CMB to higher energies due to their bulk motion. The power in the lepton beam is then constrained by the requirement that the bulk Compton (BC) emission should not overproduce the upper limits of the bridge emission.
Furthermore,  the cold lepton power  constraints can be used in the context of the EC/EMB model for the non-thermal knot emission to constrain the jet matter content:  a knot  configuration that successfully produces the knots SED will be  incompatible with an electron-positron jet if the lepton power required to be fed into the knot exceeds the BC upper limit on  the power  carried by leptons in the bridge.  This was applied by \cite{uchiyama05}
on PKS 0637-752 using Spitzer observations to argue that, in the context of the EC/CMB model, the three jet configurations they considered  are incompatible with an electron-positron jet, because the power required to be fed into the knot exceeds the upper limit of the power carried by leptons. 
  
\begin{figure}
\epsscale{0.7}
\plotone{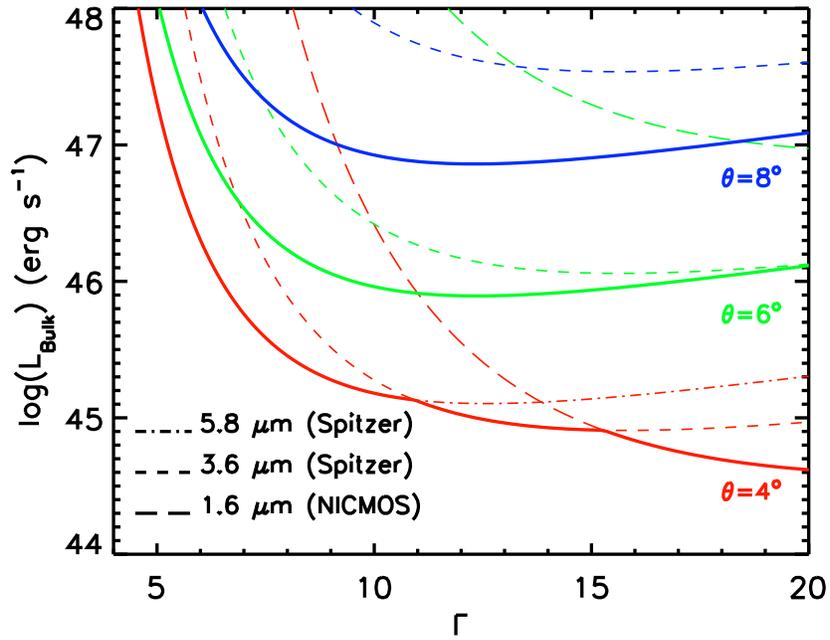}
\caption{  The solid lines are upper limits  on the bulk power of  cold leptons in the jet for three different angles, as a function of bulk Lorentz factor $\Gamma$.   These solid lines are the lower envelope  of constraints derived from the upper limits on the flux of knot WK 5.7 through 
Spitzer (Uchiyama et al. 2005) observations at $5.8 \mu ,$ (dot-dash lines), $3.6 \mu m$ (long dash line), and NICMOS  observations at $1.6 \mu m$ (short dash line).} 
\label{fig:bulk}
\end{figure}

  We calculate here the model independent constraints for the cold lepton power, applying the formalism of Georganopoulos et al. (2005) to our  $1.6 \mu $m NICMOS data of WK 5.7, as well as the Spitzer data 
  at $3.6 \, \mu $m and $5.8\,  \mu $m (as it turns out, shorter wavelengths limits  do not constrain the power further as long as the bulk Lorentz factor $\Gamma\gtrsim 20$).  
   A beam of cold electrons of  power $L_e$   that propagates a length $l$ with a bulk Lorentz factor $\Gamma$  and velocity $u=\beta c$ through  a blackbody photon field characterized by a temperature $T$,  produces a specific luminosity
\begin{equation}
L_{\nu}=\frac{L_{e}\sigma_{\rm T}l kT}{m_ec^5\beta^2\Gamma^3}\; \nu^2 \;
 \ln  \left[ \frac{1-\exp[-h\nu\Gamma(1+\beta)/(\delta\, kT)]}
         {1-\exp[-h\nu /(\Gamma \delta \, kT(1+\beta))]} \right], \label{app.lum}
\end{equation}
where $\delta=1/(\Gamma(1-\beta \cos \theta))$ and $\theta$ is the angle between the beam and the line of sight. Selecting the knot WK 5.7, as Uchiyama et al. (2005), and setting the length to $l=13.8/\sin\theta$ Kpc, we require that $L_\nu$ is smaller than the WK 5.7 knot upper limits and calculate the upper limit for $L_e$. The results are plotted 
in Figure \ref{fig:bulk}  for three different angles.  (for $\theta<4^{\circ}$ the jet becomes longer 
than 1 Mpc; $\theta\gtrsim 8^{\circ}$ is in disagreement with the superluminal motions observed in the VLBI core, unless the jet is significantly bend),  and for a range of $\Gamma$.  We see that for a given angle, for   $\Gamma \gtrsim 10$, the constraints are relatively flat with increasing $\Gamma$, going from  from $\sim 10^{45} $ erg s$^{-1}$ for $\theta=4^{\circ}$ to $\sim 10^{47} $ erg s$^{-1}$ for $\theta=8^{\circ}$.  If we assume a black hole mass of $10^9$ M$_\odot$, this suggests that for all plausible angles the cold lepton power has to be sub-Eddington
as long as $\Gamma\gtrsim 10$. We discuss the implications of these constraints on the EC/CMB model
in \S \ref{eccmb}.

\subsection{Why the magnetic field cannot dominate in the  EC/CMB model \label{noB}}

We now turn to the more constrained EC/CMB model,
 focusing our attention on  knot WK 7.8, which we model as a sphere of radius $R= 1 $ kpc (corresponding to an angular diameter of $0.3$ arcsec) permeated by a magnetic field $B$. We assume that the EED, at least in the regime responsible for the synchrotron radio and EC/CMB X-rays, is a power law $N(\gamma)=k\gamma^{-p}$,
where $N(\gamma)d\gamma$ is the total number of electrons in the source with Lorentz factors
in the range $\gamma,\;\gamma+d\gamma$. Using the $\delta$-function, energy-conserving approximations for the synchrotron and IC emissivities, the EC/CMB and synchrotron observed specific luminosity ($L_{EC}$ and $L_S$ respectively) can be written as
\begin{equation}
L_{EC}(\nu)=c_1 k \delta^{4+2\alpha}\nu^{-\alpha}, \;\; L_{S}(\nu)=c_2 k B^{\alpha+1}\delta^{3+\alpha}\nu^{-\alpha}, 
\label{eq:lslec}
\end{equation}  
where $\alpha=2p+1$, $\delta$ is the usual Doppler factor and $c_1, \; c_2$ are listed in the Appendix. The magnetic field energy density is a fraction $f$ of the radiating lepton energy density
\begin{equation}
{B^2 \over 8 \pi}=f\;{3 k m_e c^2 \gamma_{min}^{1-2\alpha}\over 4\pi R^3},
\label{eq:lb}
\end{equation}
where $\gamma_{m in}$ is the low energy cutoff of the EED and, in agreement   with the radio and X-ray observations, we have assumed $p>2$.  

\begin{figure}
\epsscale{0.7}

\plotone{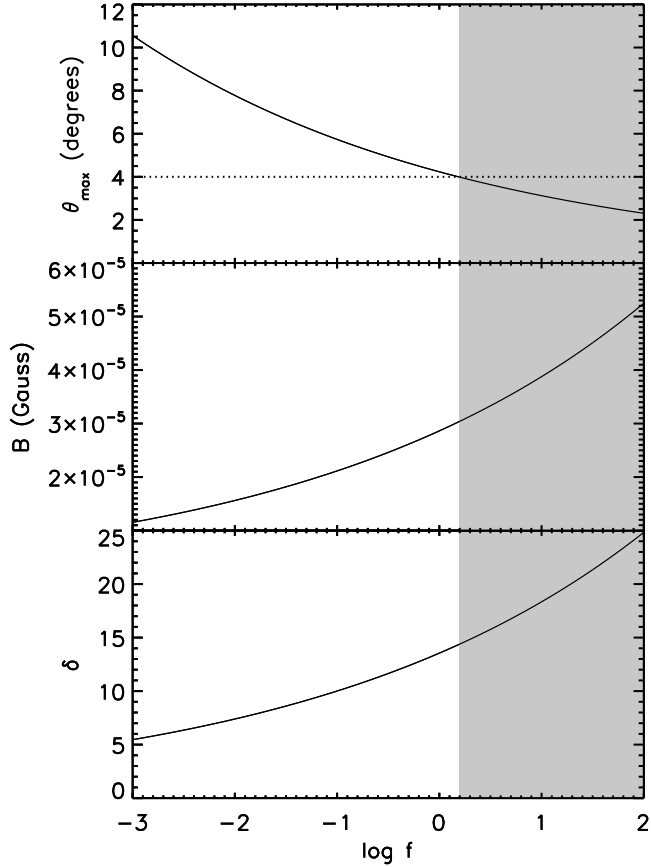}
\caption{ The Doppler factor $\delta$ (bottom panel), the magnetic field $B$ (middle panel) and the maximum permitted angle $\theta_{max}$ as a function of $f=U_B/U_{part}$, the magnetic field to radiating particle energy density. The shaded area corresponds to $\theta_{max}<4^\circ$ that requires
jets longer than $1 $Mpc. In this calculation,
we assumed $\gamma_{min}=20$. This can vary by at most $\pm 50\%$ without violating the optical and X-ray constraints (higher values underproduce the X-rays, lower values overproduce the optical).
Such variations move the curves somewhat but do not change our results significantly.
  } 
\label{fig:f}
\end{figure}

Although equipartition ($f=1$) is widely adopted 
in the study of jets, there is no physical argument for why the plasma should be in equipartition. Here, and in the rest of this work,  instead of presuming equipartition, we let $f$ be a free parameter  and we examine how it can be  constrained.  
 For a given value of $f$ the above three equations can be solved to provide $B, \; \delta, $ and $k$. For a given resulting $\delta$, the maximum angle $\theta_{max}$ that the jet can form with the line of sight is  $\sin^{-1} (1/\delta)$. We constrain the jet orientation angle to $4^\circ< \theta<9^{\circ}$, with the lower limit set by the requirement that the jet up to knot WK 9.7 is less that 1 Mpc long, and  the upper limit set by  the observed superluminal core speeds \citep{lovell00}.   As can be seen in Figure \ref{fig:f}, magnetically dominated jets
($f \gtrsim 1$) are excluded, because they require $\theta_{max}< 4^{\circ}$.  Jets in which the radiating lepton energy density is increasingly higher than that of the magnetic field require gradually  smaller $\delta$ and  larger $\theta_{max}$.  
%Note that  a jet   close to equipartition,  requires $\theta\approx 4^\circ$, $B\approx 5 \times 10^{-5}$ G, and $\delta\approx 15$. 

\subsection{ Matter content in the EC/CMB model \label{eccmb}} 

%{\bf elaborate on g+-, also mantion that some limited leptonic is ok in the cross of the lines at theta=8 deg}

\begin{figure}
\epsscale{0.7}

\plotone{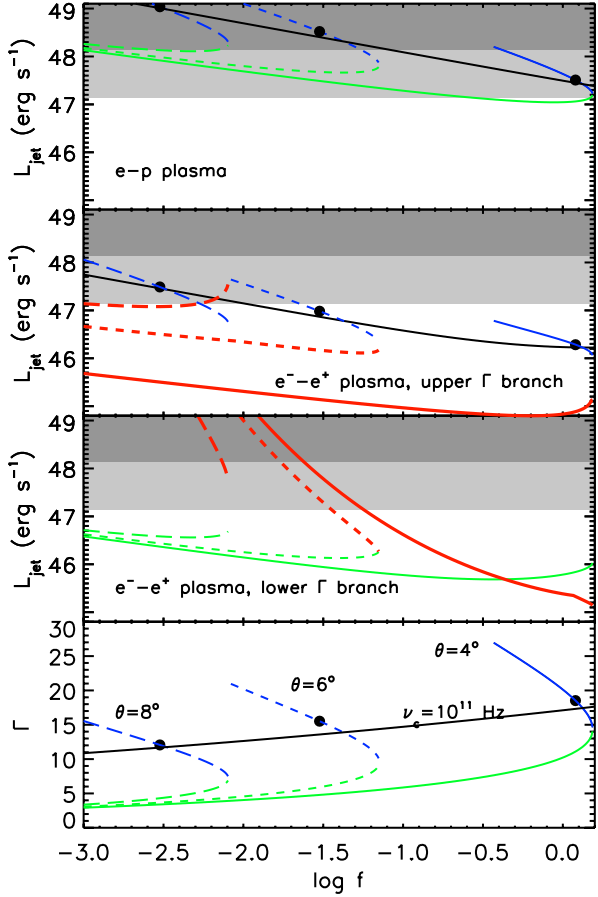}
\caption{  Bottom panel: for three different  angles, we plot, as a function of $f$,  the lower  (green) and upper (blue)  Lorentz factors $\Gamma$ that are are compatible with the radio and X-ray observations of knot WK 7.8 in the EC/CMB model. We also plot with a black line, the locus of the models that have a synchrotron cooling break frequency $\nu_c=10^{11}$ Hz. The black line has the same meaning in the top two panels. The filled circles correspond to numerical models we discuss in \S \ref{sed}. Second from the bottom panel: the  power required for $e^--e^+$ jets (green lines) and the upper limits of the cold lepton jet power  from the BC constraints (red lines), both 
 in the lower $\Gamma$-branch configurations. The shaded areas correspond to $L_{jet} > L_{Edd,9}$
 (light gray shade) and $L_{jet} > L_{Edd,10}$
 (heavy gray shade). Second from the top panel:  the  power required for $e^--e^+$ jets (blue lines) and the upper limits of the cold lepton jet power  from the BC constraints (red lines), both 
 in the upper  $\Gamma$-branch configurations. Shaded areas in this and the next panel are the  same
 as in the previous panel. Top panel: The jet power in the case of an $e-p$ jet for the lower (green) and upper (blue) $\Gamma$ branches. The three different line styles refer to the  three different angles.
 The same  style of the lines is used in all plots for each angle.  
 } 
\label{fig:edd_constr}
\end{figure}

We study here the constraints imposed on the plasma composition and Lorentz factor of the jet by the requirement that any viable configuration should reproduce the SED of knot WK 7.8 (hereafter simply called the knot) and should not be substantially super-Eddington.
Given that the plasma is not magnetically  dominated, we only examine  solutions with $f \lesssim 1$.
For a given jet orientation angle $\theta$, there are two possible values of the bulk Lorentz factor $\Gamma$ compatible with a particular $\delta$:  
\begin{equation}
\Gamma_{1,2}={ 1\pm (1-(1-\cos^2 \theta)(1+\cos^2\theta \delta^2))^{1/2} \over \delta(1-\cos^2\theta)}.
\label{eq:gamma}
\end{equation}
Selecting three representative angles ($\theta=4^\circ, 6^\circ, 8^\circ$), we let $f$ vary, and for each $f$ we calculate $\delta$ and $B$. For each value  of $\delta$ we calculate the two possible values of $\Gamma$. A constraint on $\Gamma$ can be imposed by  the
requirement that the electrons responsible for the radio emission escape  from the knot before they have time to cool,  as suggested   by the overall SED of the knots and from the fact that the radio spectrum of the knots has a spectral index $\alpha_r\approx 0.8$ that corresponds to an  EED electron index of $p=2\alpha_r+1\approx 2.6$, typical of uncooled electrons in radio sources. The requirement that  cooling does not affect the radio frequencies can be written as $\nu_c >10^{10}$ Hz, where the cooling frequency is
\begin{equation}
\nu_c={e \over 2\pi m_e}  \left[ 9 m_e c^2 \over 16 \,\sigma_\tau  k\,  R \, U_{CMB}  (1+z)^4\,  \Gamma^2 \right]^2 {B \delta \over 1+z }, 
\label{cooling}
\end{equation}
with $k$ being the escape time from the source in units of light crossing time ($k=3$ is in agreement with the jump conditions at  highly relativistic shock; Blandford \& McKees 1977), $U_{CMB}$  the CMB energy density, 
$\sigma_{\tau}$  the Thomson cross section and $z$ the redshift of the source (this equation does not take into account synchrotron losses that turn out to be negligible compared to the EC/CMB ones, as can be ssen by the fact that the X-ray luminosity is higher than the radio-optical). 
We plot the two different $\Gamma$-branches for each angle at the lower panel of Figure \ref{fig:edd_constr}, taking into account the above  constraint, which, for the range of $f$ we consider,  only affects the fast $\Gamma$ branch 
for $\theta=4^\circ, 6^\circ$  (without this constraint, these upper branches would continue toward higher values of $\Gamma$ and lower values of $f$. 
 As can be seen, if we drop the equipartition assumption,
a wide range, $3\lesssim \Gamma\lesssim 28$, can reproduce the X-ray and radio  data. 

An additional strong constraint comes from the requirement that  the SED breaks from the radio slope 
at $\sim 10^{11}$ Hz, peaks at $\sim 10^{12}$ Hz, and curves  down at IR, optical and UV bands (Uchiyama et al. 2005; see also \S \ref{sed}).  The most plausible explanation for this break is radiative cooling. Using equation (\ref{cooling}) for $\nu_c=10^{11}$ Hz,
we plot with a solid black line  in the lower panel of Figure \ref{fig:edd_constr} the locus of configurations with $\nu_c=10^{11}$ Hz. This line intersects  the upper $\Gamma$ branch for all angles. Numerical models that fit the knot SED should cluster around this line (see \S \ref{sed}).
 If, however, we choose to interpret the SED not as due to radiative cooling, but rather as due to an injected EED that is intrinsically curved, then all the configurations below the black line are possible.

We now turn to the question of the power required to feed the knot, noting that power requirements
significantly 
higher than the Eddington luminosity of a $\sim 10^9$ M$_\odot$ black hole  ($L_{Edd, 9}=1.38 \times 10^{47}$ erg s$^{-1}$) are unpleasant, and power requirements
higher than the Eddington luminosity of a $\sim 10^{10}$ M$_\odot$ black hole are disfavored. 
The power in radiating leptons and magnetic field required to feed the knots is
\begin{equation}
L_{e^--e^+}=\pi R^2 \beta c \Gamma^2 U_B(1+1/f).
\label{eq:lept}
\end{equation}
In the second from the bottom panel of Figure \ref{fig:edd_constr}, we plot (green lines) the jet power required to be fed into the knot  in the case of an $e^--e^+$ jet composition for the lower $\Gamma$-branch (as we mentioned above these configurations are plausible only if we presume that the shape of the SED is not due to
cooling, but rather due to an appropriately fine-tuned intrinsically curving EED).
 The required power is  for all cases  below $L_{Edd,9}$ (the lightly shaded area indicates $L_{jet}>L_{Edd,9}$, while the heavily shaded area  indicates $L_{jet}>L_{Edd,10}$). We also plot with red lines the upper limit to the jet power from the constraint that BC emission cannot overproduce the flux limits of knot WK 5.7. As can be seen, the only part of the parameter space that a leptonic configuration  is unattainable  (red line lower than green) and therefore a proton contribution is needed, is for configurations close to equipartition.

In the second from the top  panel we plot the same quantities for  the upper  $\Gamma$-branch, that includes the more plausible configurations in which the SED shape is due to self-consistent radiative cooling (black line).
Except for the high $f$ tail  of large angle models (in the case depicted,  $\theta=8^\circ$ models with $\log f\approx -2.1$), the rest of the $e^--e^+$ models are excluded because they require more power in the knot than allowed by the upper limit set by  the BC constraints (red lines below  blue  lines). 
This means that these models are viable only if protons carry a power at least equal to the difference between that required by the knot and that carried by the cold leptons in the bridge. 
%The black line plotted is, as in the lower panel, the locus of models with $\nu_c=10^{11}$ Hz. 
Note that large angles are disfavored because  $L_{jet} \gtrsim L_{Edd,9}$.

A particular $e-p$ jet is one in  which  for every radiating lepton there is a cold proton. The jet power required in such a jet is
is
\begin{equation}
L_{e-p}=\pi R^2 \beta c \Gamma^2 U_B\left [1+{1\over f }(1+{ m_p\over m_e<\gamma>})\right] , \; 
<\gamma >={ \gamma_{min} (p-1) \over p-2}.
\end{equation}
We plot this power in the upper panel of Figure \ref{fig:edd_constr}, together with the black lines of models with $\nu_c=10^{11}$ Hz,
 and we note that the only models that do not have extreme power requirements  ($L_{jet} \lesssim L_{Edd,9}$)  are those close to equipartition (and, by necessity, at a small angle).

To sum it up, under the most plausible assumption, that the shape of the SED is due to electron cooling,
$e^--e^+$ jets are excluded. Assuming a minimal contribution of protons, just enough to power the knot,
higher angle jets are disfavored because they require $L_{jet} \gtrsim L_{Edd,9}$. Increasing the proton contribution to one proton per radiating electron,  increases the jet power, and the only configurations that have $L_{jet} \sim L_{Edd,9}$ are those at equipartition and, consequently, small angle.

\subsection{Model SEDs  \label{sed}}
\begin{figure}
\epsscale{0.7}
\plotone{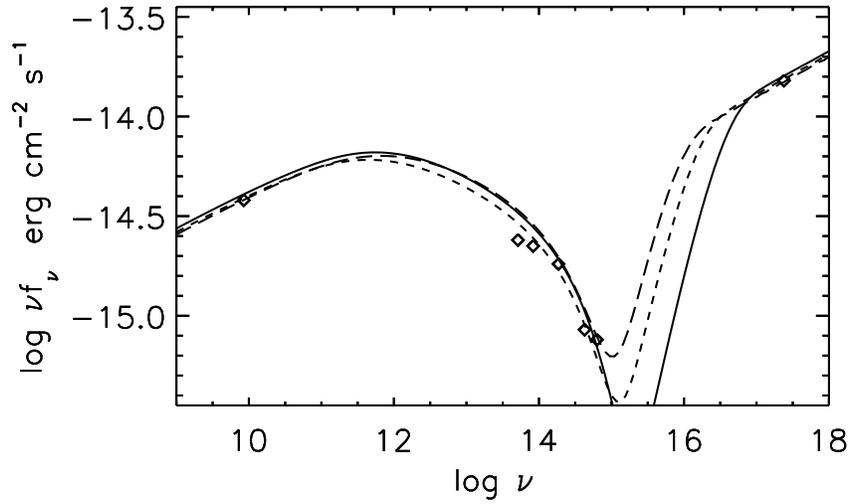}
\caption{ The data (solid points, see Table \ref{tab:fluxes}) and three model SEDs for knot WK 7.8 at 
$\theta=4^{\circ}$ (solid line), $\theta=6^{\circ}$ (short dash line), $\theta=4^{\circ}$ (long dash line). 
 In all three models, $\gamma_{min}=20$,  while   $\gamma_{max}=10^6,\, 1.5 \times 10^6,\, 2\times 10^6$   for the solid, short, and long dash lines, chosen by the requirement to fit
the optical-UV tail of the synchrotron emission.}
\label{fig:sed}
\end{figure}

To confirm that models on  or close to the black lines of Figure  \ref{fig:edd_constr}  provide reasonable representations of the  knot  SED, we present in Figure \ref{fig:sed} the results of three 
simulations\footnote{The code can be found at {\sl http://jca.umbc.edu/$\sim$markos/cs}}  at angles $\theta=4^\circ, 6^\circ, 8^\circ$. For each angle, we select a corresponding $f=1.2, 0.03, 0.003 $, and from these two parameters we calculate the rest of the model parameters from equations (\ref{eq:lslec}), (\ref{eq:lb}), (\ref{eq:gamma}), and (\ref{eq:lept}) (see Table \ref{tab:models}).
We also plot in Figure  \ref{fig:edd_constr} with solid circles the location of the models. All three models provide adequate representations of the SED and are, therefore, equally acceptable. As expected, all three models are located close to the black line of  $\nu_c=10^{11}$ Hz. Also, all three models require protons to carry some ($\theta=8^\circ$)
or most ($\theta=4^\circ, 6^\circ$) of the power.  Only the model  close to  equipartition ($\theta=4^\circ$), however, can have one proton per radiating particle and still be sub-Eddington.

\subsection{A high energy faster spine and a low energy slower sheath \label{spine}} 

Having exhausted the constraints that can be imposed on the jet physics in the context of the EC/CMB model, we now present a rather model independent argument that the jet of PKS 0637-752 is characterized by a fast spine emitting most of the IR to X-ray emission, and a slow sheath emitting most of the radio.
The first direct observational argument that a radio jes  may be characterized by a fast spine and a slower sheath came from radio observations of the several-kpc  radio jets of the FR II radio galaxy 3C 353 by  \cite{swain98}, who interpreted the apparently lower emissivity along the jet axis as Doppler de-beamed radiation from a spine with plasma flow velocities faster than those of the sheath. Similar arguments, based on radio observations of the kpc jet of the FR I radio galaxy 3C 264 and on the  
pc jet of the FR I radio galaxy M87, were presented by \cite{lara04} and \cite{kovalev07} respectively.  
A  comprehensive multiwavelength study of the Kpc scale jet of M87  \citep{perlman01, perlman05} 
shows stratification in the high energy radiation production, in the sense that effective particle acceleration capable of producing the observed infrared to X-ray synchrotron emission of M87 is concentrated along the spine of the jet. A recent study of the jets Centaurus   A shows  similar results, in the sense that 
the X-ray spectrum of the jet knots is harder along the spine \citep{worrall08}.

A model-independent argument for a spine-sheath geometry in the jet of PKS 0637-752, with the spine both ({\sl i}) being    faster and ({\sl ii}) emitting most of the IR to X-ray emission, can be constructed on the basis of multiwavelength observations and on the assumption that the quasar has two jets of similar power. This assumption is supported by the fact that in radio galaxies and quasars,  the level of the presumably non-beamed  radio emission from the two lobes is comparable 
(e.g.  Perley, R\"oser, Meisenheimer 1997).
 The morphology of the western jet strongly suggests that the X-ray knots are part of the jet and not termination shocks. This is also supported by the fact that the western knot complex magnetic field orientation, determined by radio polarization measurements \citep{lovell00}, is parallel to the jet axis and not perpendicular, as one would expect from strong perpendicular shocks, typically found at the terminal hot spots.
Radio observations \citep{lovell00} show a radio feature diametrically opposite and at approximately the same distance from the core as the complex of the three knots WK 7.8, WK 8.9, and WK 9.7. 
 Interestingly, the same parallel to the jet axis magnetic field orientation is observed at the eastern jet radio feature, suggesting that this is also not a terminal hotspot, and that the jet plasma flows through the radio feature. 
 
  Assuming that the two jets are intrinsically symmetric, the ratio $j_r\approx1.5$ of the jet to counter jet radio fluxes (see Figure \ref{fig:counterjet} and Table \ref{tab:fluxes}) constrains the radio plasma speed $u_r=\beta_r c$ to be
\begin{equation}
\beta_r \cos\theta={j_r^{1/(3+a)}-1\over j_r^{1/(3+a)}+1}\approx 0.06,
\end{equation}
 where $a=0.8$ is the radio spectral index, and $\theta$ the jet orientation to the line of sight. If the IR to X-ray emitting plasma had the same velocity 
as the radio, one would expect a clear detection of the counter jet in the IR to X-ray range. However, this is not the case and using the limit $j_X\gtrsim 100$ for the jet to counter jet X-ray ratio one obtains $\beta_X cos\theta\gtrsim 0.54$. The speed, therefore, of the X-ray emitting plasma must be significantly higher than that of the radio plasma.
Similar, but somewhat lower speed constraints can be derived using the IR and optical ratio limits.

\section{Conclusions \label{conclusions}}

We present new NICMOS and ACS observations, which we use, together with existing broadband observations to build  the most complete SED of the jet knots of quasar PKS 0637-752. 
We relax the equipartition assumption and  show that in the EC/CMB framework for the X-ray emission, the plausible requirement that the $10''$ jet  is at most 
1 Mpc long excludes magnetically dominated jets.
 We also show that limits on the BC emission exclude
$e^--e^+$ jets in the EC/CMB framework, regardless of equipartition: the jet needs to carry some, or most of its power with protons. Interestingly, if we require one proton per radiation lepton, only the equipartition jet does not require substantially super-Eddington power.  
Taking radiative cooling into account, the Lorentz factor $\Gamma\sim 18-20$ required to  model the knot SED for the equipartition solution is significantly higher that those used previously, and  is more in agreement with
the radio core limit $\Gamma > 17.4$, derived  from core superluminal motions \citep{lovell00}. 
Configurations with significantly more than one protons per radiating lepton are excluded in the EC/CMB model.
Finally, we note that the counter jet radio feature and its polarization, together with the limits on the IR-optical and X-ray emission, suggest a spine sheath flow, with most of the IR-optical-X-ray emission
coming from the spine. 

\acknowledgements

Part of this work was done in the context of the senior thesis of Kushal Mehta  at UMBC.
The authors acknowledge support from LTSA grants \# NAG5-9997 and NNG05GD63G at UMBC, as well as \#NNX07AM17G at FIT, and from the  HST observing grant GO-10541.01  at UMBC.
%\section {Conclusions \label{conc}}

\appendix

\section{The parameters $c_1, \, c_2$}

The parameters $c_1$  and $ c_2$, using a $\delta$-function approximation for the synchrotron and IC
emissivity are given by
\begin{equation}
c_1={   4^{(p-2)/2} \sigma_{\tau}c  U  \over   3^{(p-1)/2}  }   \left( {h\over K_BT}\right)  ^{(3-p)/2} ,\;\;\;
c_2={\sigma_{\tau} c B_{crit}^{(3-p)/2}\over 12 \pi }\left( {h\over m_e c^2}\right)  ^{(3-p)/2},
\end{equation}
where $U$and $T$  are the CMB photon energy density and temperature respectively  at the redshift of the source, $\sigma_{\tau}$ is the Thomson cross section, and $B_{crit}=(m_e^2 c^3) /(e \hbar)$ is the critical magnetic field.

\clearpage
\begin{deluxetable}{llllllll}
\tabletypesize{\scriptsize}
\rotate
\tablecaption{Flux Densities of the Various Parts of PKS 0637-752. The errors in the
NICMOS and ACS fluxes are 5\% of the flux value. The radio data are taken
from~\cite{lovell00}, the Spitzer  from~\cite{uchiyama05} and the X-rays
from~\cite{chartas00} and \cite{schwartz00}.}
\tablewidth{0pt}
\tablecolumns{8}
\tablehead{
\colhead{Telescope} &
\colhead{Frequency (Hz)} & \colhead{WK 5.7} & \colhead{WK 7.8} & \colhead{WK 8.9} 
& \colhead{ WK 9.7} & \colhead{Total Western Jet$^*$} & \colhead{Counter jet feature} \\
}
\startdata
ATCA & $8.60\times 10^9$ & $1.60\times 10^{-15}$ & $3.78\times 10^{-15}$ & $3.31\times
10^{-15}$ & $3.72\times 10^{-15}$ & $1.12\times 10^{-14}$
 & $8.60\times 10^{-15}$ \\
Spitzer & $5.17\times 10^{13}$ & $2.59\times 10^{-15}_L$ & $2.38\times 10^{-15}$ & $3.31\times
10^{-15}$ & \nodata & $1.07\times 10^{-14}$ 
 & \nodata \\
Spitzer  & $8.33\times 10^{13}$ & $1.67\times 10^{-15}_L$ & $2.25\times 10^{-15}$ & $3.17\times
10^{-15}$ & \nodata & $8.75\times 10^{-15}$ 
 & \nodata \\
HST/NICMOS & $1.87\times 10^{14}$ & $2.81\times 10^{-16}$ & $1.83\times 10^{-15}$ & $1.96\times
10^{-15}$ & $7.99\times 10^{-16}$ & $4.88\times 10^{-15}$
 & $1.68\times 10^{-15}_L$ \\
HST/WFPC2 & $4.30\times 10^{14}$ & $4.00\times 10^{-16}_L$ & $8.61\times 10^{-16}$ & $1.21\times
10^{-15}$ & $4.05\times 10^{-16}$ & $2.87\times 10^{-15}$
 & \nodata \\
HST/ACS & $6.32\times 10^{14}$ & $9.21\times 10^{-16}_L$ & $7.54\times 10^{-16}$ & $8.77\times
10^{-16}$ & $4.05\times 10^{-16}$ & $2.96\times 10^{-15}$
 & $1.69\times 10^{-16}_L$ \\
Chandra & $2.42\times 10^{17}$ & $2.42\times 10^{-15}$ & $1.52\times 10^{-14}$ & $1.77\times
10^{-14}$ & $1.02\times 10^{-14}$ & $4.55\times 10^{-14}$
 & $4.84\times 10^{-16}_L$ \\
\enddata
\tablenotetext{a}{The fluxes given here are in units of $ergs cm^{-2} s^{-1}$}
\tablenotetext{*}{Total Western Jet$ = WK 5.7 + WK 7.8 + WK 8.9 + WK 9.7$}
\tablenotetext{L}{Upper Limit}
\label{tab:fluxes}
\end{deluxetable}

%------------------------------------------------------------

\clearpage
\begin{deluxetable}{ccccccc}
\tablecaption{Parameters for the SED models. $\theta$ and $f$ are the free parameters.}
\tablewidth{0pt}
\tablecolumns{7}
\tablehead{
\colhead{$\theta$} & \colhead{$f$} & \colhead{$B$} & \colhead{$\delta$} 
& \colhead{ $\Gamma$} & \colhead{$L_{e^--e^+}$} & \colhead{$L_{e-p}$ } \\
 &  & \colhead{$(10^{-5}G)$} &   &  & \colhead{ (erg s$^{-1}$)} & \colhead{ (erg s$^{-1}$)}}
\startdata
$4^\circ$ & $1.2$ & $2.9$ & $13.9$ & $18.5$ & $1.9 \times 10^{46}$
 & $3.2 \times 10^{47}$ \\ 
 $6^\circ$ & $0.03$ & $1.8$ & $8.5$ & $15.5$ & $9.6 \times 10^{46}$
 & $3.3 \times 10^{48}$ \\ 
 $8^\circ$ & $0.003$ & $1.3$ & $6.3$ & $12.7$ & $3.1 \times 10^{47}$
 & $1.1 \times 10^{49}$ \\ 
 \enddata
\label{tab:models}
\end{deluxetable}

\clearpage

%------------------------------------------------------------

%------------------------------------------------------------

%------------------------------------------------------------

%------------------------------------------------------------

%------------------------------------------------------------

%------------------------------------------------------------

\end{document}